\documentclass[pdflatex, sn-nature, iicol]{sn-jnl}

\setcitestyle{numbers,super,open={},close={}}

\usepackage{xr-hyper}
\usepackage{graphicx}%
\usepackage{multirow}%
\usepackage{amsmath,amssymb,amsfonts}%
\usepackage{amsthm}%
\usepackage{mathrsfs}%
\usepackage[title]{appendix}%
\usepackage{xcolor}%
\usepackage{textcomp}%
\usepackage{manyfoot}%
\usepackage{booktabs}%
\usepackage{algorithm}%
\usepackage{algorithmicx}%
\usepackage{algpseudocode}%
\usepackage{listings}%
\usepackage{makecell}
\usepackage{multicol}
\usepackage{mathtools}
\usepackage{tikz}
\usepackage{booktabs}
\usepackage{subfig}
\usepackage{booktabs}
\usepackage{array}
\usepackage{siunitx}
\usepackage{float}
\usepackage{placeins}
\usepackage{subfiles}

\usepackage[switch]{lineno}

\theoremstyle{thmstyleone}%
\theoremstyle{thmstyletwo}%

\theoremstyle{thmstylethree}%

\begin{document}


\title[Article Title]{Multiconfiguration Pair-Density Functional Theory Calculations of Low-lying States of Complex Chemical Systems with Quantum Computers}


\author[1]{\fnm{Zhanou} \sur{Liu}}

\author[3]{\fnm{Yuhao} \sur{Chen}}

\author[6,7]{\fnm{Yingjin} \sur{Ma}} 

\author*[3,4,5]{\fnm{Xiao} \sur{He}}\email{xiaohe@phy.ecnu.edu.cn}

\author*[2]{\fnm{Yuxin} \sur{Deng}}\email{yxdeng@msg.sufe.edu.cn}

\affil[1]{\orgdiv{Shanghai Key Laboratory of Trustworthy Computing}, \orgname{East China Normal University}, \orgaddress{\street{} \city{Shanghai}, \postcode{200062}, \state{} \country{China}}}

\affil[2]{\orgdiv{MoE Key Laboratory of Interdisciplinary
Research of Computation and Economics}, \orgname{Shanghai University of Finance and Economics}, \orgaddress{\street{} \city{Shanghai}, \postcode{200433}, \state{} \country{China}}}

\affil[3]{\orgdiv{Shanghai Engineering Research Center of Molecular Therapeutics and New Drug Development, Shanghai Frontiers Science Center of Molecule Intelligent Syntheses, School of Chemistry and Molecular Engineering}, \orgname{East China Normal University}, \orgaddress{\street{} \city{Shanghai}, \postcode{200062}, \state{} \country{China}}}

\affil[4]{\orgdiv{Chongqing Key Laboratory of Precision Optics}, \orgname{Chongqing Institute of East China Normal University}, \orgaddress{\street{} \city{Chongqing}, \postcode{401120}, \state{} \country{China}}}

\affil[5]{\orgdiv{New York University–East China Normal University Center for Computational Chemistry}, \orgname{New York University Shanghai}, \orgaddress{\street{} \city{Shanghai}, \postcode{200262}, \state{} \country{China}}}

\affil[6]{\orgdiv{Computer Network Information Center}, \orgname{Chinese
 Academy of Sciences}, \orgaddress{\street{} \city{Beijing}, \postcode{100190}, \state{} \country{China}}}

 \affil[7]{\orgdiv{National Super-Computing Center in CAS}, \orgname{Chinese
 Academy of Sciences}, \orgaddress{\street{} \city{Beijing}, \postcode{100190}, \state{} \country{China}}}

\abstract{Accurately describing strong electron correlation in complex systems remains a prominent challenge in computational chemistry as near-term quantum algorithms treating total correlation often require prohibitively deep circuits. Here we present a hybrid strategy combining the Variational Quantum Eigensolver with Multiconfiguration Pair-Density Functional Theory to efficiently decouple correlation effects. This approach confines static correlation to a compact multireference quantum state while recovering dynamic correlation through a classical on-top density functional using reduced-density information. By enabling self-consistent orbital optimization, the method significantly reduces quantum resource overheads without sacrificing physical rigor. We demonstrate chemical accuracy on standard benchmarks by reproducing C$_2$ equilibrium bond lengths and benzene excitation energies with mean absolute errors of 0.006 Å and 0.048 eV respectively. Most notably, for the strongly correlated Cr$_2$ dimer requiring a large complete active space (48e, 42o), the framework yields a bound potential-energy curve and recovers qualitative dissociation behavior despite realistic hardware noise. These results establish that separating correlation types provides a practical route to reliable predictions on near-term quantum hardware.
}




\maketitle

\section*{Introduction}\label{sec1}

The simulation of molecular electronic structures lies at the heart of quantum chemistry, enabling predictions of chemical reactivity, spectroscopic properties, and thermodynamic behaviors. Accurate resolution of molecular electronic structures necessitates the simultaneous and adequate treatment of both static and dynamic electron correlations. Conventional Hartree-Fock (HF)~\cite{hartree1928wave, slater1929theory, slater1930note} theory falls short in capturing these two types of correlations. In response, classical computational approaches have developed several valuable improvements, which can be broadly categorized into two groups, multireference methods represented by configuration interaction (CI)~\cite{roothaan1951new}, and single-reference methods, such as coupled cluster (CC)~\cite{kummel2003biography, bartlett2007coupled}, perturbation theory~\cite{cremer2011moller}, and density functional theory (DFT)~\cite{hohenberg1964inhomogeneous, levy1979universal, vignale1987density}.

However, solving the electronic Schr\"{o}dinger equation for realistic molecular systems remains a formidable challenge due to the exponential scaling of computational resources with system size. Quantum computing, with its inherent parallelism and ability to represent entangled quantum states~\cite{nielsen2010quantum}, offers a transformative path forward. The variational quantum eigensolver (VQE) serves as a fundamental framework for quantum computation, leveraging hybrid quantum-classical optimization to accommodate the limited coherence times of near-term hardware. Early work in quantum computational chemistry employed VQE to reproduce classical methods such as HF~\cite{google2020hartree} or unitary coupled-cluster singles and doubles (UCCSD)~\cite{lee2018generalized}. While effective for dynamic correlation, these single-reference approaches often lack a complete description of static correlation. To address this, recent studies~\cite{parrish2019quantum, greene2021generalized, mizukami2020orbital, sokolov2020quantum, takeshita2020increasing, yalouz2021state, sugisaki2022variational, khan2023chemically, fitzpatrick2024self, zhao2024quantum} have adopted multireference ansätze inspired by the complete active space self-consistent field (CASSCF) method~\cite{roos1987complete}.

Despite this progress, when combined with CI-like expansions, UCCSD-style circuits can become prohibitively large, motivating the search for alternative routes to recover dynamic correlation. Furthermore, the practical realization of quantum computational chemistry on noisy intermediate-scale quantum (NISQ)~\cite{bharti2022noisy} devices remains constrained by hardware imperfections, limited qubit connectivity, and algorithmic inefficiencies.

To overcome these limitations, there is a need for alternatives to UCCSD that retain accurate dynamic correlation while reducing circuit depth. Multiconfiguration density functional theory (MC-PDFT)~\cite{li2014multiconfiguration} integrates concepts from density functional theory into the configuration interaction framework, achieving a balanced description of both static and dynamic correlation at relatively low computational cost. In this approach, static correlation is handled by the multireference wavefunction, while the remaining short-range dynamic correlations are efficiently processed using a classical on-top pair-density functional. MC-PDFT functionals utilize both the density $\rho(\text{r})$ and the on-top pair density $\Pi(\text{r})$---the probability of finding two electrons at the same point in space. This approach is specifically designed to minimize the double-counting of electron correlation. Inspired by MC-PDFT, we have designed a corresponding quantum circuit capable of representing multireference molecular wave functions expanded in large basis sets.

The central focus of this study is the introduction of VQE-MC-PDFT, a hybrid quantum-classical algorithm that integrates VQE with MC-PDFT. This methodology adopts a strategic approach to tackle the electron correlation challenge. Initially, the quantum segment, VQE, addresses the static correlations within an active space that bears chemical significance, which presents difficulties for conventional methods. This resolution is accomplished through a streamlined multireference quantum circuit. Subsequently, the remaining dynamic correlations are efficiently processed using a classical density functional, which is incorporated into the quantum computation framework. This integrated strategy aims to deliver precise results while reducing demands on near-term quantum hardware, particularly concerning the circuit complexity.

Building on these considerations, Fig.~\ref{Overview-S} summarizes the proposed VQE-MC-PDFT workflow; technical details are provided in the Supplementary Information. In contrast to the UCCSD ansatz
\begin{equation}
|\Psi(\boldsymbol{\theta})\rangle
=\exp\!\big[\hat T(\boldsymbol{\theta})-\hat T^{\dagger}(\boldsymbol{\theta})\big]\,
|\Psi_{0}\rangle,  
\label{eq.UCCSD}
\end{equation}
where the truncated cluster operator \(\hat T\) comprises single- and double-excitation terms from occupied to virtual orbitals, our scheme uses VQE to prepare a multiconfigurational wave function in a chemically motivated active space and extracts the associated one- and two-electron reduced density matrices (RDMs) via tailored measurement schedules. Direct UCCSD implementations on quantum hardware incur a controlled-NOT (CNOT) count scaling as \(\mathcal{O}\!\big(N(N-\eta)^{2}\eta^{2}\big)\) with the number of spin-orbitals \(N\) and electrons \(\eta\), which often exceeds near-term hardware limits.

In MC-PDFT, dynamic correlation is recovered by evaluating an on-top pair-density functional \(E_{\mathrm{ot}}[\rho(\mathbf r),\Pi(\mathbf r)]\) from the RDMs of a multireference state, avoiding the need for deep single-reference correlation circuits. In this work, we adopt the fully-translated PBE (ftPBE) on-top functional. The resulting energy expression, together with its contribution to the orbital-rotation gradient, is used to update orbital rotations self-consistently within the VQE-MC-PDFT loop. This separation of static and dynamic correlation leads to substantially shallower circuits with fewer entangling gates while retaining accuracy for strongly correlated systems.

Recent VQE-PDFT work~\cite{D5SC07528A} established that a VQE-prepared multiconfigurational state can be augmented by a pair-density functional correction evaluated from measured RDMs. Relative to that baseline, the present work focuses on practical feasibility and scalability in the near-term regime by introducing three extensions: (i) a self-consistent orbital-optimization procedure driven by the ftPBE-based orbital-rotation gradient, (ii) a circuit-partitioning strategy that decomposes weakly entangled clusters and enables larger active spaces via classical reconstruction, and (iii) a density-informed readout-error mitigation protocol that improves RDM and energy accuracy under measurement noise. We validate the workflow experimentally on the Tianji--S2 13-qubit superconducting device, and we employ circuit partitioning for larger active-space calculations, including Cr$_2$, reaching an effective scale corresponding to 84 qubits through fragment reconstruction, rather than a monolithic execution on hardware. To the best of our knowledge, these results constitute among the largest digital Cr$_2$ simulations reported within partitioned VQE-style workflows.

\section*{Results}\label{sec2}

\begin{figure*}[htbp!]
\centerline{\includegraphics[width=\textwidth]{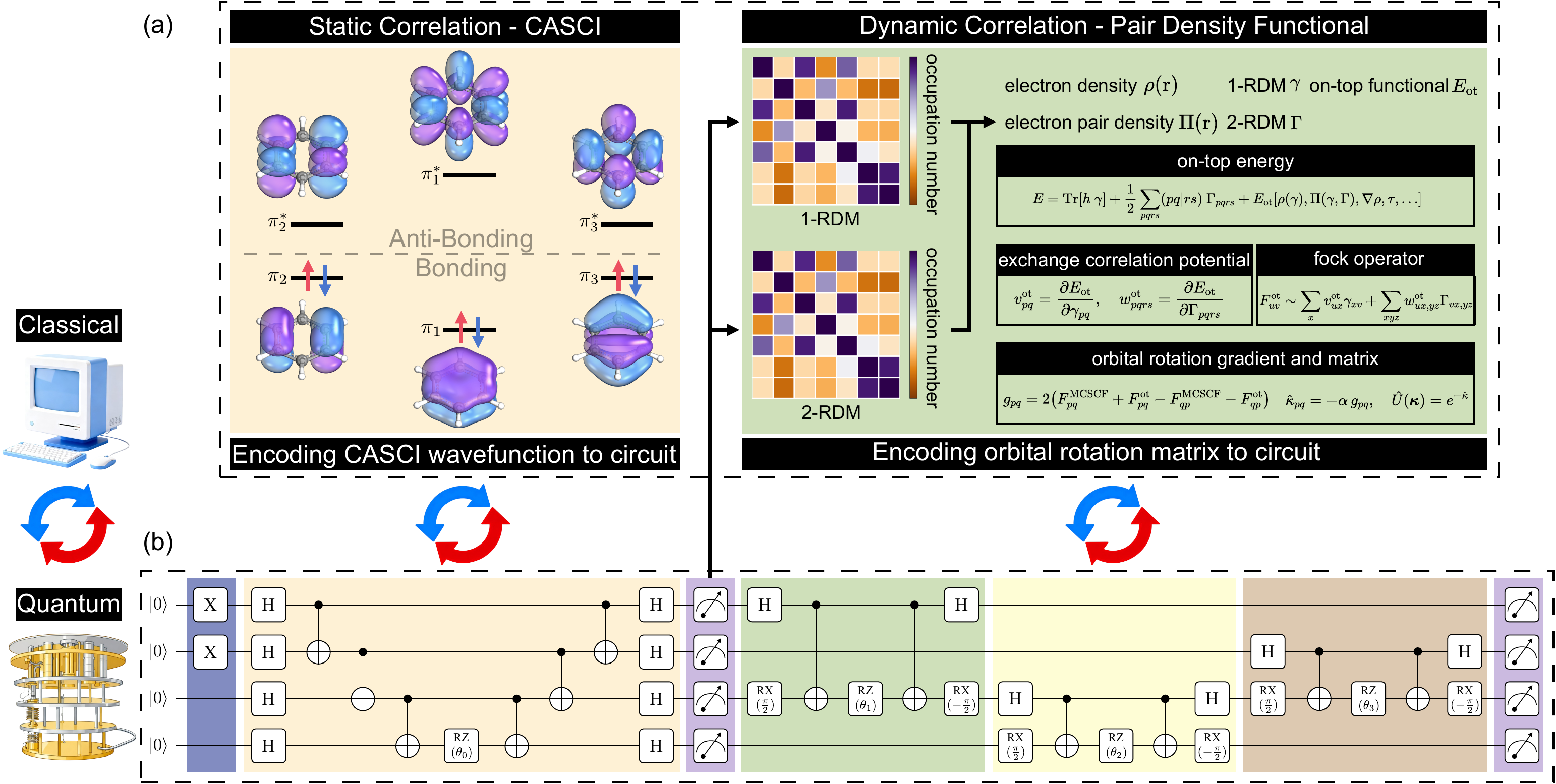}}
\caption{\textbf{Schematic of the VQE-MC-PDFT hybrid quantum-classical workflow.} \textbf{a} Algorithmic framework for decoupling electron correlation. Static correlation is treated within a chemically motivated active space (illustrated by the benzene $\pi$ system) using a CASCI wavefunction, while dynamic correlation is recovered via a classical on-top pair-density functional. Measured one- and two-body reduced density matrices (1-RDM and 2-RDM) serve as inputs to compute the on-top energy $E_{\text{ot}}$ and the orbital rotation generator $\hat{\kappa}$. \textbf{b} Quantum circuit implementation. The circuit architecture prepares the multireference ansatz (beige block) followed by a parameterized orbital rotation unitary $\hat{U}(\boldsymbol{\kappa})$ (green, yellow, and brown blocks). This rotation stage is obtained by compiling the classical orbital-rotation generator $\boldsymbol{\kappa}$ from the MC-PDFT orbital gradient, enabling the self-consistent variational optimization of both configuration coefficients and molecular orbitals.}
\label{Overview-S}
\end{figure*}

\begin{figure}[ht!]
\centering
\includegraphics[width=9.1cm]{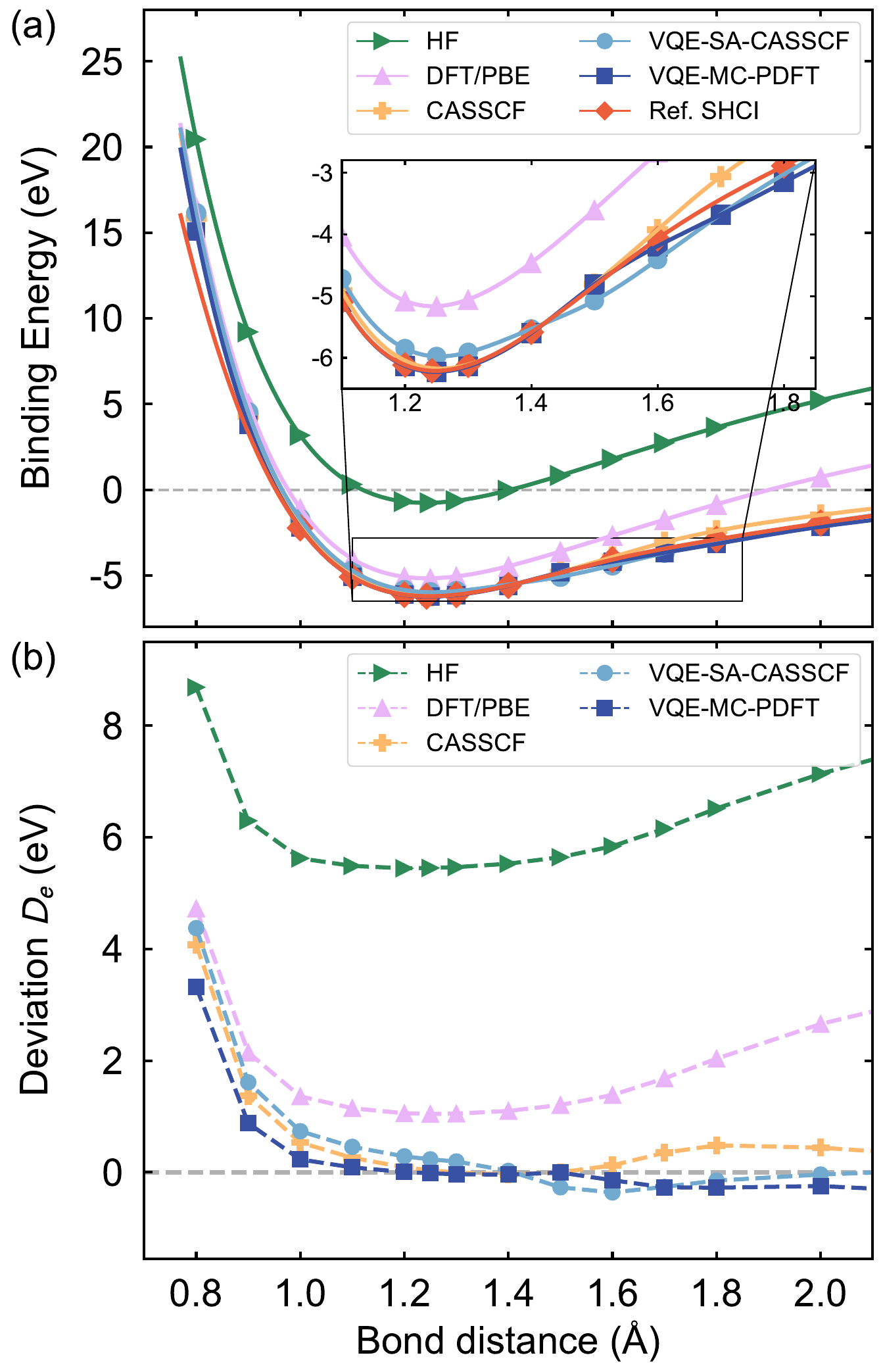}
\caption{
\textbf{Ground-state binding energy curves and error analysis for the carbon dimer ($\text{C}_2$).} \textbf{a} Potential energy curves (PECs) computed using VQE-MC-PDFT compared with Hartree-Fock (HF), DFT (PBE functional), CASSCF, and VQE-SA-CASSCF methods. The reference curve (red diamonds) is from high-accuracy Semistochastic Heat-bath Configuration Interaction (SHCI) calculations. The inset highlights the region near the equilibrium bond length. \textbf{b} Energy deviation ($D_e$) in the binding energy curves relative to the SHCI benchmark along the bond coordinates. For comparison on the bond-length grid used here, the SHCI values were spline-interpolated onto the plotted bond lengths.}
\label{ground-c2}
\end{figure}

The elemental simplicity of the C$_2$ molecule belies a deceptively intricate electronic structure. Owing to the energetic proximity of several electron configurations, a valid description of most of its low-energy electronic states requires more than a single dominant configuration~\cite{jiang2022diabatic}. This multireference character is a defining feature of its electronic ground state $X^1\Sigma_g^+$.
To verify whether our VQE-MC-PDFT method can accurately describe the electronic structure of this strongly correlated molecular system, we calculated the potential energy curves (PECs) of the C$_2$ ground state $X^1\Sigma_g^+$ using the VQE-MC-PDFT method, and compared with other quantum chemical methods.

The top panel of Fig.~\ref{ground-c2} illustrates the PECs from various quantum chemical methods for the dissociation of the ground state of the C$_2$ molecule. The bottom panel provides a clearer comparison by plotting the energy deviation ($D_e$) of each method’s binding-energy curve relative to the SHCI benchmark~\cite{holmes2017excited} (cc-pVQZ), where the SHCI data were spline-interpolated onto the bond-length grid used in this work and no constant energy shift was applied.


The multireference nature of the C$_2$ molecule during bond dissociation necessitates methods that go beyond a single-reference description. Standard quantum chemistry approaches such as HF and DFT yield qualitatively incorrect results, such as a purely repulsive potential energy curve. 
CASSCF can provide qualitatively correct PECs; however, it systematically underestimates the binding energy due to the neglect of dynamic correlation effects.
On this foundation, we evaluate both the variational quantum eigensolver-based state-averaged CASSCF (VQE-SA-CASSCF)~\cite{zhao2024quantum} and our proposed VQE-MC-PDFT.

Benchmarking against SHCI data shows close agreement of VQE-MC-PDFT across the PECs and $D_{e}$ deviations.
It accurately reproduces the SHCI potential energy surface with high precision, in stark contrast to VQE-SA-CASSCF, which merely captures the qualitative trends of its classical counterpart. The minimal and consistent deviations of VQE-MC-PDFT from the zero-error line further underscore its high accuracy.


The enhanced accuracy is most evident near the equilibrium bond distance of 1.25~\AA. At this point, the VQE-MC-PDFT binding energy of -6.227~eV agrees excellently with the SHCI reference value of -6.209~eV. In contrast, the VQE-SA-CASSCF result of -5.977~eV deviates significantly. This performance underscores VQE-MC-PDFT's balanced treatment of both static and dynamic electron correlation.

Furthermore, the VQE-MC-PDFT curve exhibits a smoother and more consistent profile. The deviation plot highlights a noticeable fluctuation in the VQE-SA-CASSCF curve in the intermediate region around 1.50~\AA. This is likely an artifact of hardware noise that is notably absent in the VQE-MC-PDFT result, demonstrating our method's robustness under the same experimental conditions. 

\begin{figure}[t!]
\centering
\includegraphics[width=\columnwidth]{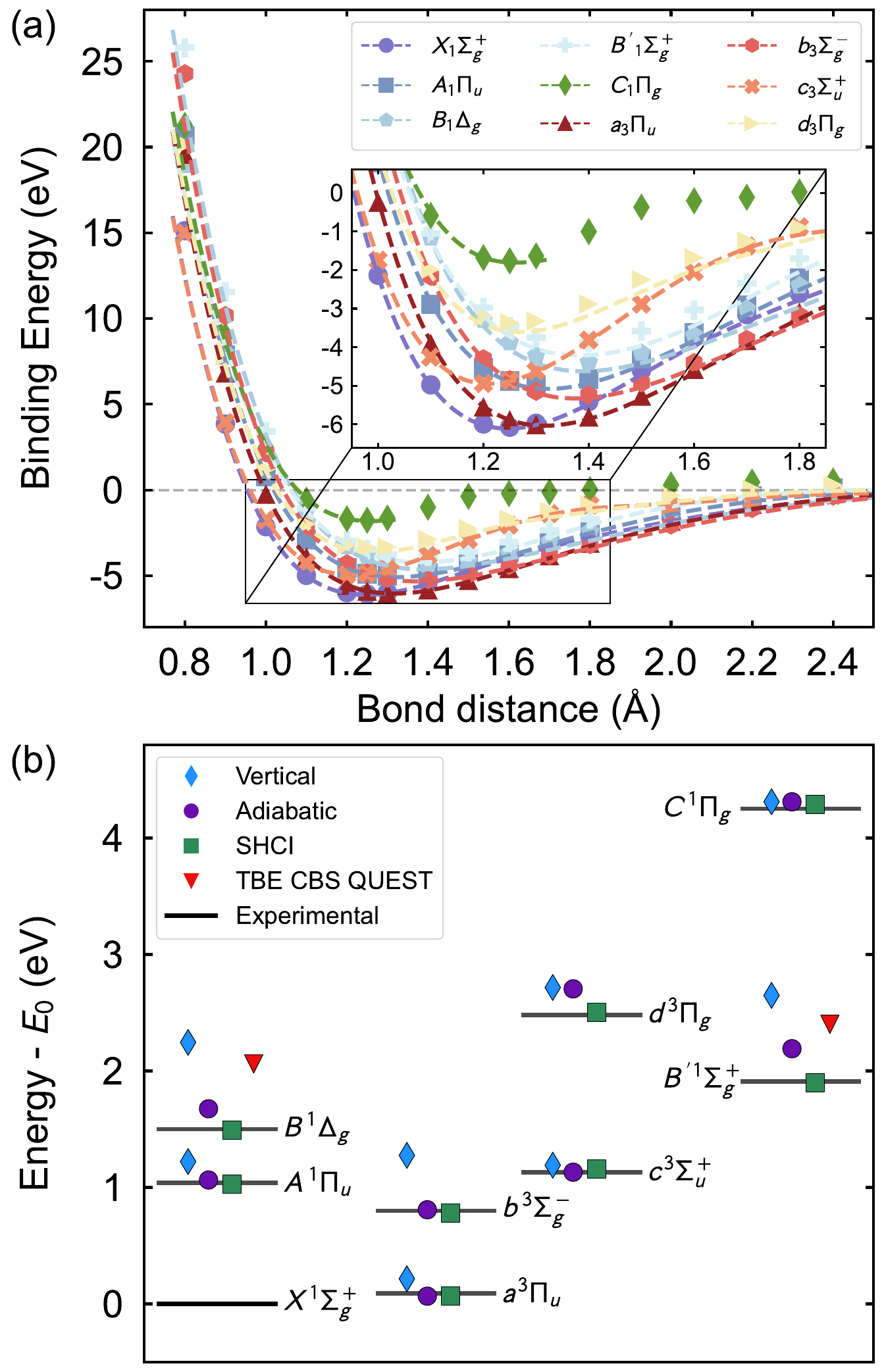}

\caption{
\textbf{Low-lying electronic structure of the carbon dimer (C$_2$) from VQE-MC-PDFT.} \textbf{a} Potential energy curves for the ground state ($X^1\Sigma_g^+$) and eight lowest-lying singlet and triplet excited states. Dashed lines represent the VQE-MC-PDFT results, while markers indicate the SHCI reference data. The inset provides a magnified view of the equilibrium region, illustrating the method's ability to resolve dense manifolds of states and their crossings. \textbf{b} Comparison of calculated vertical (blue diamonds) and adiabatic (purple circles) excitation energies against experimental values (black lines), SHCI data (green squares), and theoretical best estimates (TBE) from the QUEST database (red triangles). Our calculations use the cc-pVTZ basis, whereas the SHCI reference data were reported in the cc-pV5Z basis; therefore, the SHCI PECs shown here are shifted upward by a constant offset of +0.108~Ha (2.94~eV) to align the dissociation limit on the cc-pVTZ energy scale.
}
\label{excited-c2}
\end{figure}


We applied the VQE-MC-PDFT method to compute the PECs of the ground state ($X^1\Sigma_g^+$) and the eight lowest-lying singlet and triplet excited states of the C$_2$ molecule. The resulting curves, shown in the top panel of Fig.~\ref{excited-c2}, are smooth and physically well-behaved over a wide range of internuclear distances, confirming the robustness of our approach from the equilibrium region to dissociation. These curves not only capture the characteristic potential wells essential for deriving spectroscopic constants but also accurately reproduce the conical intersection present in the reference SHCI results. The SHCI reference data~\cite{holmes2017excited} were obtained at the cc-pV5Z level and spline-interpolated onto the bond-length grid used in this work; to align the dissociation limits for visual comparison on the cc-pVTZ energy scale, a constant offset of $+0.108$~Ha ($\approx 2.94$~eV) was applied to the SHCI PECs. This constant shift cancels out in the calculation of excitation energies and therefore does not affect the comparisons presented in the bottom panel. Therefore, the key quantitative benchmarks are the energy differences---specifically, the vertical and adiabatic excitation energies, which are compared in the bottom panel of the figure against experimental data and theoretical best estimates (TBE) from the QUEST database~\cite{veril2021questdb} of highly accurate excitation energies.

\begin{table}[t!]
\centering
\caption{\textbf{Comparison of calculated and experimental spectroscopic constants for the carbon dimer (C$_2$).}
The table lists equilibrium bond lengths ($R_{eq}$ in \AA) and adiabatic excitation energies (in eV) for the ground state and eight excited states. VQE-MC-PDFT results are compared against experimental values (Exp)~\cite{martin1992c2}. The Mean Absolute Error (MAE) demonstrates the method's accuracy in reproducing geometric and energetic properties.}
\label{tab:c2_spectroscopy}
\renewcommand{\arraystretch}{1.2}
\begin{tabular*}{\columnwidth}{ p{0.12\columnwidth}<{\centering} p{0.16\columnwidth}<{\centering} p{0.16\columnwidth}<{\centering} p{0.16\columnwidth}<{\centering} p{0.16\columnwidth}<{\centering}}
\toprule
\multirow{2}{*}{State} & \multicolumn{2}{c}{$R_\text{eq}$(\AA)} & \multicolumn{2}{c}{Excitation energy (eV)} \\
\cmidrule(r){2-3} \cmidrule(r){4-5}
 & This work & Exp & This work & Exp \\
\hline

$X^1\Sigma_g^+$ & 1.24484 & 1.24244   & 0.00 & 0.00    \\
$a^3\Pi_u$      & 1.31091 & 1.31188  & 0.07 & 0.09 \\
$b^3\Sigma_g^-$ & 1.36597 & 1.36924  & 0.81 & 0.80 \\
$A^1\Pi_u$      & 1.31311 & 1.31831  & 1.06 & 1.04 \\
$c^3\Sigma_u^+$ & 1.20740 & 1.20937  & 1.13 & 1.13 \\
$B^1\Delta_g$   & 1.38138 & 1.38547 & 1.68 & 1.50 \\
$B'^1\Sigma_g^+$& 1.37257 & 1.37735 & 2.19 & 1.91 \\
$d^3\Pi_g$      & 1.26246 & 1.26610  & 2.70 & 2.48 \\
$C^1\Pi_g$      & 1.24264 & 1.25516  & 4.31 & 4.25 \\
\midrule
MAE & 0.006 & -- & 0.10 & -- \\
\bottomrule
\end{tabular*}
\end{table}

The quantitative performance of our VQE-MC-PDFT method is summarized in Table~\ref{tab:c2_spectroscopy}, which compares our calculated spectroscopic constants with established experimental values~\cite{martin1992c2}. The method demonstrates exceptional accuracy in predicting molecular geometries: for all nine states, the calculated equilibrium bond lengths (\(R_\text{eq}\)) exhibit a mean absolute error (MAE) of only 0.006~\AA\ relative to experiment. This precision is maintained for adiabatic excitation energies, where the VQE-MC-PDFT results achieve an MAE of 0.10~eV. Such accuracy, approaching experimental fidelity, marks a significant step for a near-term quantum algorithm and a substantial improvement over classical CASSCF. 

Performance is particularly notable for the lower-lying states. For instance, the excitation energies for the \(A^1\Pi_u\) and \(c^3\Sigma_u^+\) states deviate from experiment by merely 0.02~eV and 0.00~eV, respectively. While overall agreement is excellent, larger deviations are observed for the \(B^1\Delta_g\) (0.18~eV) and \(B'^1\Sigma_g^+\) (0.28~eV) states. These states are characterized by significant double-excitation character, which is notoriously challenging for most quantum chemical methods to capture. The ability of VQE-MC-PDFT to yield both qualitatively correct energy ordering and reasonable quantitative estimates for such demanding states strongly underscores its advanced capability.

\begin{figure}[t!]
\centering
\includegraphics[width=\columnwidth]{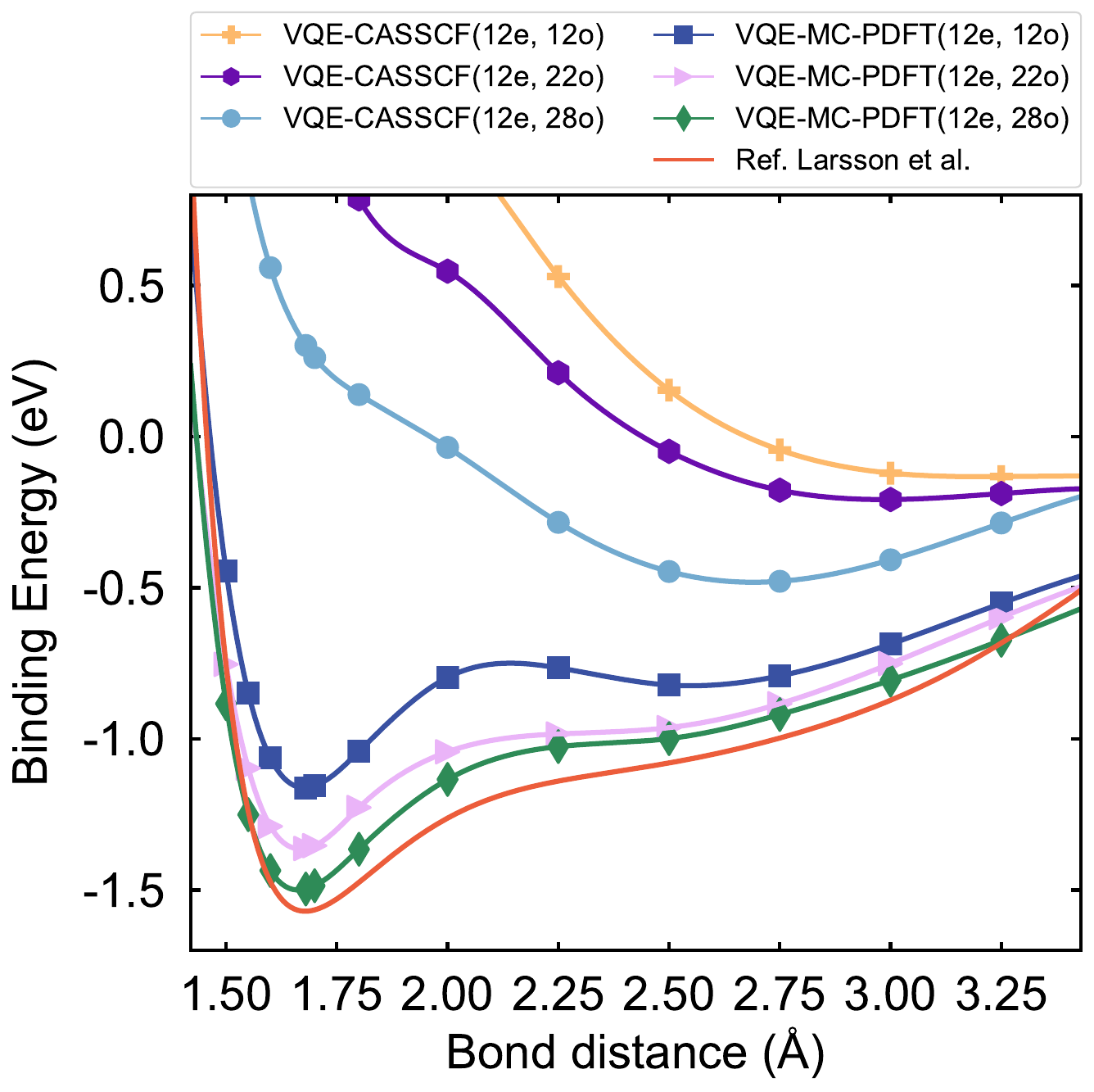}


\caption{\textbf{Active-space scaling evaluation of VQE-CASSCF and VQE-MC-PDFT potential energy curves for Cr$_2$.}
Potential energy curves (PECs) computed using Variational Quantum Eigensolver Complete Active Space Self-Consistent Field (VQE-CASSCF) and VQE Multiconfiguration Pair-Density Functional Theory (VQE-MC-PDFT). Results are compared for three active spaces: CAS(12e, 12o), CAS(12e, 22o), and CAS(12e, 28o), using the cc-pVTZ-DK basis set. The VQE-MC-PDFT method (solid markers) yields bound potentials even in minimal active spaces, whereas VQE-CASSCF (open markers) predicts qualitatively incorrect repulsive curves. The reference curve (red line; Ref. Larsson et al.~\cite{larsson2022chromium}) is included for comparison.}

\label{ground-cr2-12o-22o-ccpvtzdk}
\end{figure}

\begin{figure}[htbp]
\centering
\includegraphics[width=\columnwidth]{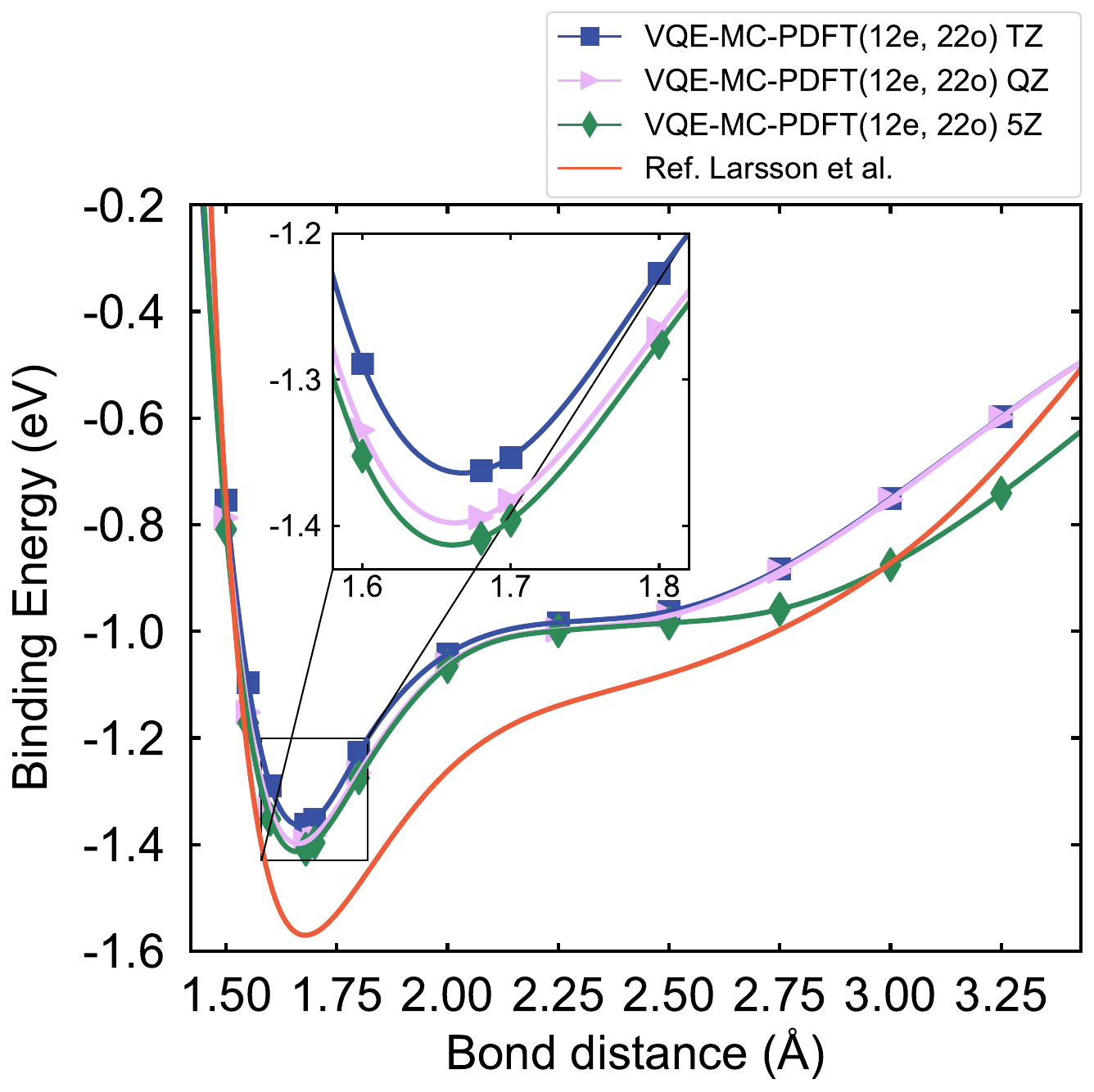}



\caption{\textbf{Basis set convergence of VQE-MC-PDFT potential energy curves for Cr$_2$.}
Calculations employ the CAS(12e, 22o) active space with correlation-consistent polarized valence X-zeta Douglas-Kroll basis sets (cc-pVXZ-DK), where X = T (Triple), Q (Quadruple), and 5 (Pentuple). The inset magnifies the region near the equilibrium bond length to illustrate convergence. The reference curve (Ref. Larsson et al.~\cite{larsson2022chromium}) is included for comparison. Extending the basis set from Triple- to Pentuple-zeta yields consistent well depths, indicating that residual errors are dominated by static correlation within the active space rather than the one-electron basis set size.}

\label{ground-cr2-12o-22o-ccpwcvtzdk}
\end{figure}

The PECs of the chromium dimer present a formidable challenge for quantum chemical methods~\cite{kurashige2011second}. On the one hand, the molecule features a formal sextuple bond, and the minimum active space required for correct dissociation is CAS(12e, 12o). Therefore, static correlation effects are crucial, as evidenced by the qualitatively incorrect repulsive curve predicted by standard single-reference methods, such as CCSD(T)~\cite{bauschlicher1994cr2}. On the other hand, dynamic correlation effects are equally significant, as even using the density matrix renormalization group (DMRG) to solve the large CAS(48e, 42o) active space still fails to provide an accurate potential energy curve for Cr$_2$~\cite{li2020accurate}.

In Fig.~\ref{ground-cr2-12o-22o-ccpvtzdk}, we compare VQE-CASSCF and VQE-MC-PDFT potential energy curves for Cr$_2$ as a function of the active-space size, using active spaces (12e, 12o), (12e, 22o), and (12e, 28o). The CASSCF curves remain qualitatively incorrect across these choices: the (12e, 12o) and (12e, 22o) results are predominantly repulsive over the investigated bond-length range, and even the enlarged (12e, 28o) space yields only a shallow minimum at an overly long bond distance relative to the Larsson reference. By contrast, VQE-MC-PDFT produces a bound potential already in the minimal valence space and improves systematically upon active-space expansion. Increasing the active space from 12o to 22o and further to 28o deepens the well and reduces the deviation from the reference curve across the near-equilibrium and intermediate-distance regions, indicating that residual errors are dominated by incomplete valence correlation within the chosen active space. 

To assess basis-set convergence at fixed active space, Fig.~\ref{ground-cr2-12o-22o-ccpwcvtzdk} reports VQE-MC-PDFT results for the (12e,22o) active space using cc-pVXZ-DK basis sets with X = T, Q, and 5 (TZ/QZ/5Z). The VQE-MC-PDFT curves for TZ/QZ/5Z are closely clustered over the bond-length range, demonstrating that the basis-set dependence is comparatively modest for this active-space choice. Moving from TZ to QZ and 5Z yields only small changes in the well depth and the deviation profile (bottom panel), with the largest visible differences occurring away from the minimum (longer bond distances). Overall, the data indicate that, within the current framework, enhancing agreement with the reference curve is more effectively accomplished by expanding the active space rather than by further increasing the one-electron basis from TZ to QZ/5Z.

\begin{figure}[htbp!]
\centering
\includegraphics[width=\columnwidth]{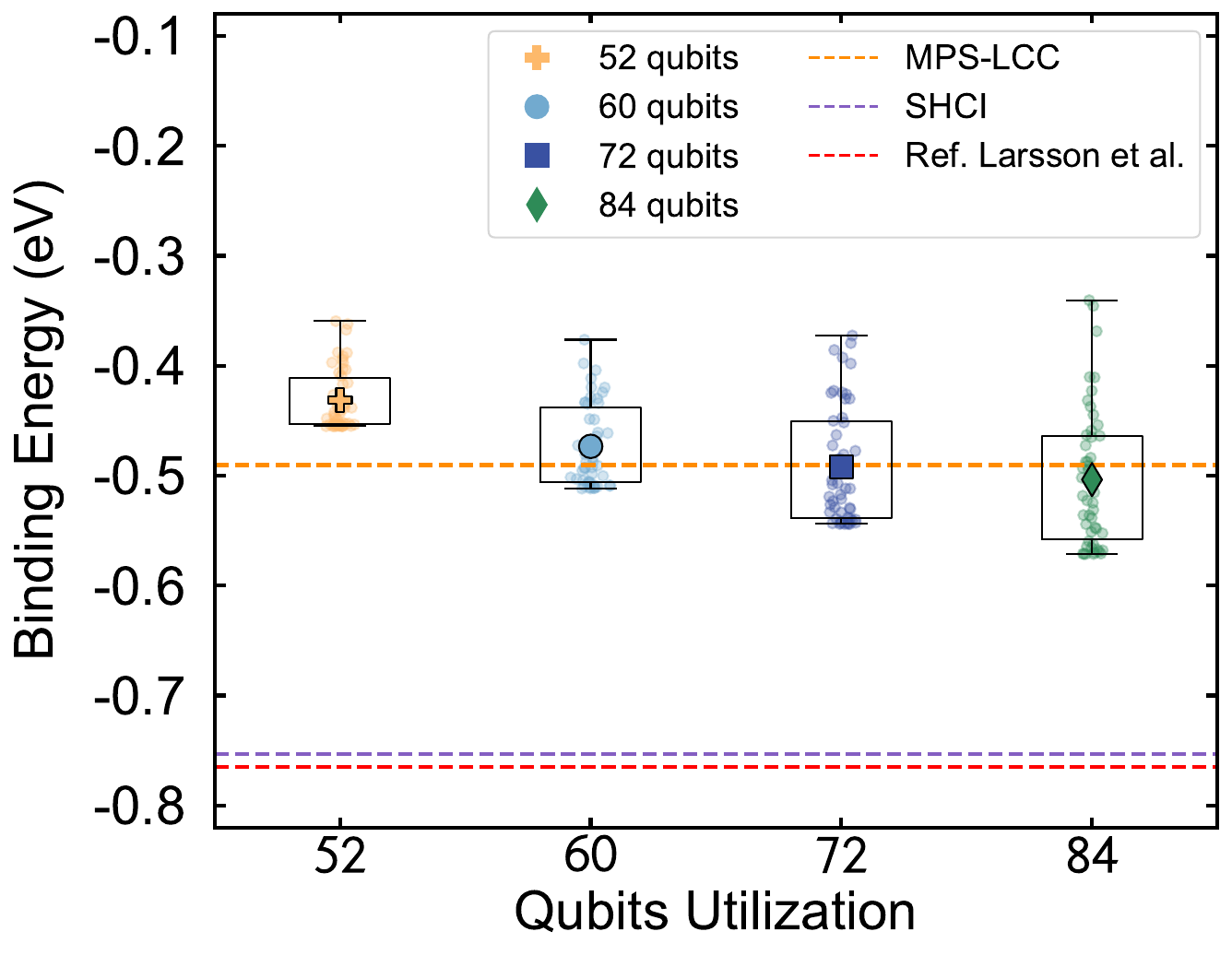}

\caption{\textbf{Scalability and convergence of Cr$_2$ binding energy distributions with quantum resource utilization.} 
Box plots showing the distribution of binding energies at a bond length of $R = 1.50$~\AA, computed using error-mitigated VQE-MC-PDFT. 
The x-axis represents the computational scale, ranging from 52 to 84 qubits (corresponding to CAS(48e, 42o)). Statistical definitions: For each qubit setting, data are aggregated from $n=50$ independent runs. The central diamond markers indicate the mean values. The box limits represent the interquartile range (IQR), spanning from the 25th to the 75th percentiles. The whiskers extend to the minimum and maximum values outside the quartiles. Colored circles represent individual data points with random horizontal jitter for visibility. Results are compared against classical benchmarks: Matrix Product State Linearized Coupled Cluster (MPS-LCC, dashed orange)~\cite{sharma2015multireference}, Semistochastic Heat-bath Configuration Interaction (SHCI, dashed blue)~\cite{li2020accurate}, and Ref. Larsson et al. (dashed red)~\cite{larsson2022chromium}.}

\label{ground-cr2-scale}
\end{figure}

The chromium dimer, which exhibits an experimental equilibrium bond length approximately 1.68~\AA, has been extensively studied at the shorter 1.50~\AA\ geometry, which serves as a stringent benchmark. Although the Ahlrichs-SV basis set used in this benchmark is considered modest for Cr$_2$, this specific geometry-basis combination has become a canonical testbed for evaluating the accuracy and scalability of high-level electronic structure methods. Seminal studies employing Multireference Linearized Coupled Cluster theory using Matrix Product States (MPS-LCC)~\cite{sharma2015multireference} and  SHCI~\cite{li2020accurate} have 
established approximate reference energies of full configuration interaction (FCI) quality for the large (48e, 42o) active space in this setting, providing the critical benchmarks against which we validate our approach.

Building directly on this foundation, we conducted a scalability and convergence study of our error-mitigated VQE-MC-PDFT method on the same system. On typical modern computational architectures, active spaces of around 18 electrons in 18 orbitals ($2 \times 10^9$ determinants) are tractable, while large-scale parallel supercomputers can partially converge results for systems up to 22 electrons in 22 orbitals ($5.2 \times 10^{11}$ determinants). Fig.~\ref{ground-cr2-scale} plots the computed binding-energy distributions as a function of the number of qubits used in our circuit partitioning and simulation strategy. With the scaling of quantum resources from 52 to 84 qubits, we observe clear and systematic convergence toward the established reference values, demonstrating the methodical improvability of our approach.


At the maximum simulated scale of 84 qubits, corresponding to the CAS(48e, 42o) space, we obtain a mean total energy of -2086.43710~Ha. This value lies remarkably close to the definitive classical benchmarks: -2086.43490~Ha from MPS-LCC~\cite{sharma2015multireference} and -2086.44456~Ha from SHCI~\cite{li2020accurate}. In terms of binding energy, the VQE-MC-PDFT result of -0.465~eV is in good agreement with high-level classical benchmarks, positioning it between the MPS-LCC value (-0.428~eV) and the SHCI value (-0.691~eV). Moreover, the tightening of the energy distribution in the box plots with increasing qubit count reflects a reduction in statistical uncertainty and enhanced numerical precision of the quantum simulation. By reproducing benchmark-quality energies for one of quantum chemistry's most challenging systems, this study demonstrates VQE-MC-PDFT as a robust, scalable, and accurate method for strongly correlated systems on near-term quantum hardware.

\begin{figure}[t!]
\centering
\includegraphics[width=\columnwidth]{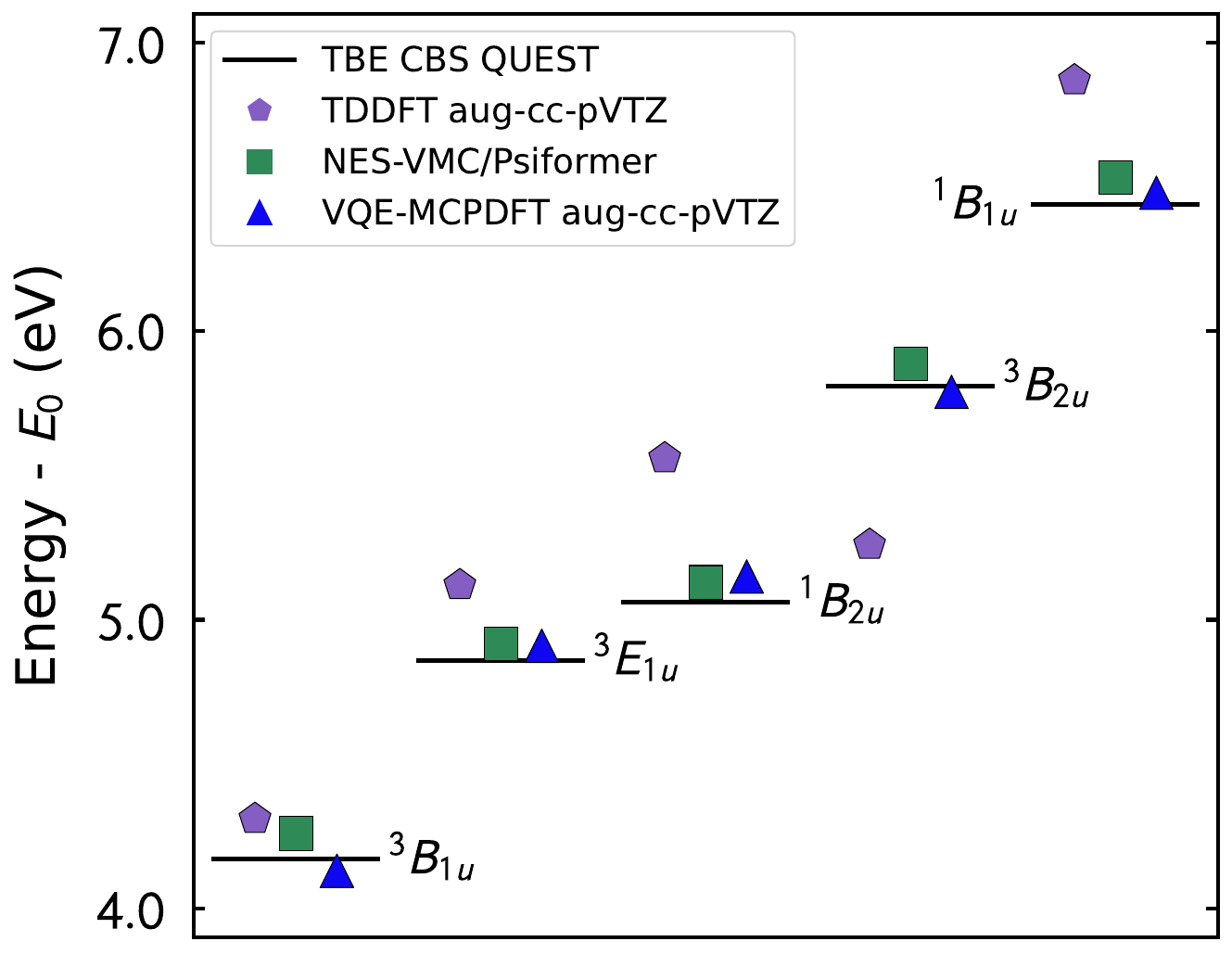}
\caption{\textbf{Vertical $\pi \to \pi^*$ excitation energies of benzene.}
Comparison of excitation energies (in eV) for the low-lying singlet and triplet states ($^3B_{1u}$, $^3E_{1u}$, $^1B_{2u}$, $^3B_{2u}$ and $^1B_{1u}$). VQE-MC-PDFT results (green triangles) are computed using a complete active space of 6 electrons in 6 orbitals (CAS(6e, 6o)), the PBE functional, and the aug-cc-pVTZ basis set. These are benchmarked against high-accuracy Theoretical Best Estimates from the QUEST database (QUEST TBE~\cite{veril2021questdb}, black lines), Time-Dependent Density Functional Theory (TDDFT, purple circles), and Natural Excited States Variational Monte Carlo (NES-VMC~\cite{pfau2024accurate}, blue squares).}

\label{excited-benzene}
\end{figure}

Finally, we applied our VQE-MC-PDFT method to the excited states of benzene, a canonical benchmark for assessing the accuracy of quantum chemical methods for aromatic systems. We compare our results, presented in Fig.~\ref{excited-benzene}, against high-level TBE from the QUEST database~\cite{benzene-TBE}, as well as standard time-dependent density functional theory (TDDFT) and the highly accurate, recently developed natural excited states variational Monte Carlo (NES-VMC) method (implemented with the Psiformer ansatz)~\cite{pfau2024accurate}.

The results show a MAE of 0.048~eV vs QUEST TBE across five $\pi$ $\to$ $\pi^*$ states, indicating high accuracy.
This level of accuracy is competitive with the state-of-the-art NES-VMC/Psiformer method (MAE of 0.076~eV) and represents a dramatic improvement over standard TDDFT with the PBE functional, which produces a significantly larger MAE of~0.37~eV and fails to correctly order the states.

The method's ability to achieve high accuracy for a chemically relevant molecule like benzene underscores the potential of the MC-PDFT framework. By effectively capturing the necessary electron correlation through its hybrid quantum-classical approach, VQE-MC-PDFT proves to be a robust and reliable tool for computing excited state properties, representing a significant step toward achieving high accuracy for challenging molecular systems on near-term quantum devices.

\section*{Discussion}
Previous work in quantum computational chemistry has often been constrained by an uneven treatment of static and dynamic correlation or by the need to operate in very small basis sets~\cite{guo2024experimental}. Here we show that VQE--MC-PDFT provides a robust and accurate pathway for simulating strongly correlated molecules on near-term quantum hardware by combining a quantum treatment of static correlation with a classical on-top pair-density functional description of the remaining dynamic correlation. Across C$_2$, Cr$_2$, and benzene, this balanced division of labor yields quantitative agreement with high-level benchmarks for both ground and excited states and offers a clear route to systematic improvement as quantum resources increase.

The C$_2$ dimer illustrates the strength of this approach in a canonical setting where bond breaking and multireference character are essential. As shown in Fig.~\ref{ground-c2} and Table~\ref{tab:c2_spectroscopy}, VQE--MC-PDFT reproduces the potential energy curve over the bond-length range investigated with high fidelity and yields spectroscopic constants with near-experimental accuracy, outperforming methods that deviate more strongly along the dissociation coordinate. These results indicate that the hybrid framework captures the interplay of static and dynamic correlation that governs both dissociation physics and multireference excited states.

The chromium dimer provides a stringent probe of the method's current scope and clarifies its dominant error channels. Our results in Figs.~\ref{ground-cr2-12o-22o-ccpvtzdk} and~\ref{ground-cr2-12o-22o-ccpwcvtzdk} show that the accuracy of VQE--MC-PDFT is predominantly driven by active-space completeness: across cc-pVXZ-DK (X = T, Q, 5), the energies change only modestly, indicating that residual errors are dominated by missing correlation within the active space rather than by the basis outside it. This behavior highlights a limitation in describing semi-core effects when they are not adequately represented in the active space, but it also points to a systematic path for improvement via active-space expansion.

We further validated scalability by studying Cr$_2$ at 1.50~\AA\ with a large (48e, 42o) active space enabled by circuit partitioning (Fig.~\ref{ground-cr2-scale}). Increasing the available qubit count from 52 to 84 yields systematic convergence toward reference energies. At the largest scale, we obtain a mean binding energy of $-0.465$~eV, highly competitive with definitive classical benchmarks such as MPS-LCC ($-0.428$~eV), while the narrowing box plots indicate reduced statistical uncertainty. These results support VQE--MC-PDFT as a robust and improvable route to benchmark-quality accuracy for a prototypical strongly correlated system.

Beyond diatomics, benzene demonstrates that the framework is effective for chemically relevant, medium-sized molecules. VQE--MC-PDFT achieves accuracy competitive with high-level classical coupled-cluster references for excitation energies (Fig.~\ref{excited-benzene}), indicating that recovering dynamic correlation through an efficient classical functional can be more effective than increasing circuit depth or ansatz complexity alone.

We note several limitations that motivate future work. Semi-core correlation remains sensitive to active-space definition, and further progress will require automated active-space selection and more efficient partitioning strategies to enable systematic active-space expansion. As in other variational algorithms, measurement costs can be substantial, motivating continued advances in measurement-efficient estimators and variance-reduction strategies for the observables entering MC-PDFT. Finally, the transferability of the on-top functional, particularly for challenging excited states, warrants further validation and potential functional development. Overall, these results suggest that separating static and dynamic correlation within a variational quantum--classical workflow offers a practical template for chemically predictive multireference simulations on near-term quantum devices.

\section*{Methods}\label{me}
\small{\subsection*{Hybrid VQE–MC-PDFT framework}}
The present work integrates a variational multiconfiguration pair-density functional theory (MC-PDFT) treatment into a hybrid quantum–classical variational quantum eigensolver (VQE) framework. Static correlation is described by a multiconfigurational active-space wavefunction encoded on a quantum processor, while dynamic correlation is incorporated through an on-top pair-density functional evaluated on a classical processor. The key ingredients are: (i) a CASCI-type ansatz implemented as a parameterized quantum circuit, (ii) measurement of active-space reduced density matrices (RDMs) from the converged VQE state, (iii) classical evaluation of the MC-PDFT energy and its functional derivatives with respect to the 1-RDM and 2-RDM, and (iv) feedback of the resulting orbital-rotation generator as a unitary circuit appended to the original ansatz. Iterating this loop yields a fully variational VQE–MC-PDFT scheme in which both configuration coefficients and orbital shapes are optimized self-consistently.

\small{\subsection*{Fermionic hamiltonian and qubit represen-\\tation}}
We consider the electronic Hamiltonian in second quantization~\cite{helgaker2014molecular},
\begin{equation}
\hat{H}
= \sum_{pq} h_{pq}\,\hat{a}^{\dagger}_{p}\hat{a}_{q}
+ \frac{1}{4}\sum_{pqrs} h_{pqrs}\,\hat{a}^{\dagger}_{p}\hat{a}^{\dagger}_{q}\hat{a}_{s}\hat{a}_{r},
\label{eq:ham-fermion}
\end{equation}
where $\hat{a}^{\dagger}_{p}$ ($\hat{a}_{p}$) creates (annihilates) an electron in spin-orbital $p$ and
$\{ \phi_p(\mathbf{x}) \}$ denotes an orthonormal spin-orbital basis.
The one-electron integrals are $h_{pq}=\langle p|\hat{h}|q\rangle$, and the two-electron integrals
$h_{pqrs}$ are taken in antisymmetrized form,
\begin{equation}
h_{pqrs} \equiv \langle pq\Vert rs\rangle
= \langle pq|rs\rangle - \langle pq|sr\rangle .
\end{equation}
Here the Coulomb integrals are defined as
\begin{equation}
\langle pq|rs\rangle
= \int \mathrm{d}\mathbf{x}_{1}\,\mathrm{d}\mathbf{x}_{2}\;
\phi^{*}_{p}(\mathbf{x}_{1})\,\phi^{*}_{q}(\mathbf{x}_{2})
\frac{e^2}{4\pi \epsilon_{0}} \frac{1}{r_{12}}\,
\phi_{r}(\mathbf{x}_{1})\,\phi_{s}(\mathbf{x}_{2}),
\end{equation}
where $\mathbf{x}=(\mathbf{r},\sigma)$ collects spatial and spin coordinates, $r_{12}=|\mathbf{r}_{1}-\mathbf{r}_{2}|$, $e$ is the elementary charge, and $\epsilon_{0}$ is the vacuum permittivity.

A fermion-to-qubit transformation such as Jordan–Wigner, parity, or Bravyi–Kitaev mapping~\cite{jordan1928paulische,seeley2012bravyi,bravyi2002fermionic} is applied to Eq.~\eqref{eq:ham-fermion}, yielding a qubit Hamiltonian of the form
\begin{equation}
\hat{H} = \sum_{j} h_{j} \hat{P}_{j} = \sum_{j} h_{j} \bigotimes_{i=1}^{N_{q}} \sigma_{i}^{(j)},
\label{eq:pauliprod-si}
\end{equation}
where each $\hat{P}_{j}$ is a Pauli string composed of single-qubit operators $\sigma_{i}^{j} \in \{I,X,Y,Z\}$, $h_{j} \in \mathbb{R}$ are scalar coefficients, and $N_{q}$ denotes the number of qubits required to encode the active spin orbitals. 
Orbital/qubit counting convention. In the second-quantized occupation representation used throughout this work, one qubit encodes one spin-orbital. Therefore, a CAS specified with $n_{\text{act}}$ spatial orbitals corresponds to $2n_{\text{act}}$ spin-orbitals and thus $N_q = 2n_{\text{act}}$ logical qubits (before any symmetry reduction).

\small{\subsection*{CASCI-inspired quantum ansatz and VQE optimization}}
The static correlation is treated within a chosen complete active space (CAS) comprising $n_{act}$ spatial orbitals and $N_{act}$ electrons. In the classical picture, a complete active space configuration interaction (CASCI) wavefunction can be written as
\begin{equation}
|\Psi_{\text{CAS}}\rangle
= \sum_{k=1}^{N_c} c_k |\Phi_k\rangle
\label{eq:casci}
\end{equation}
where $|\Phi_{k}\rangle$ are Slater determinants (or configuration state functions) spanning the active space and $c_{k}$ are configuration coefficients.

On the quantum device, each Slater determinant \(|\Phi_k\rangle\) is encoded as a computational-basis bitstring over the active qubits. We start from the Hartree--Fock (HF) reference determinant \(|\Psi_{\mathrm{HF}}\rangle\), prepared as the product state \(|1100\cdots\rangle\) that reproduces the HF occupation pattern. A parameterized, particle-number-conserving ansatz \(U_{\mathrm{ansatz}}(\vec{\theta})\) is then applied to generate a multireference trial state,
\begin{equation}
|\Psi(\vec{\theta})\rangle
= U_{\mathrm{ansatz}}(\vec{\theta}) |\Psi_{\mathrm{HF}}\rangle
= \sum_{k=1}^{N_c} c_k(\vec{\theta}) |\Phi_k\rangle.
\label{eq:ansatz}
\end{equation}
Here, the coefficients \(c_k(\vec{\theta})\) denote amplitudes in the determinant (computational) basis. Importantly, writing the energy as a sum of measured Pauli-string expectation values (see below) is a measurement decomposition of \(\hat{H}\) and does not imply that \(|\Psi(\vec{\theta})\rangle\) is separable or ``product-state encoded''.

The circuit \(U_{\mathrm{ansatz}}(\vec{\theta})\) is implemented as a sequence of one- and two-qubit gates,
\begin{equation}
U_{\mathrm{ansatz}}(\vec{\theta})
= U_M(\theta_M)\cdots U_2(\theta_2) U_1(\theta_1),
\end{equation}
with each gate chosen to preserve the total particle number (and, when required, the spin projection) while enabling exploration of the complete CASCI manifold.

Within the VQE framework~\cite{peruzzo2014variational,mcclean2016theory}, the variational energy is evaluated as
\begin{equation}
E(\vec{\theta})
= \langle\Psi(\vec{\theta})|\hat{H}|\Psi(\vec{\theta})\rangle
= \sum_j h_j \langle \Psi(\vec{\theta})|\hat{P}_j|\Psi(\vec{\theta})\rangle,
\label{eq:vqe-energy}
\end{equation}
where the Hamiltonian \(\hat{H}\) is expressed as a weighted sum of Pauli strings \(\hat{P}_j\). Each \(\langle \hat{P}_j\rangle\) is estimated from repeated measurements after rotating the relevant qubits into the measurement basis implied by \(\hat{P}_j\) (identity/\(Z\): no rotation; \(X\): apply \(H\); \(Y\): apply \(S^\dagger H\), consistent with the convention \(|y\pm\rangle=(|0\rangle\pm i|1\rangle)/\sqrt{2}\)). A classical optimizer updates \(\vec{\theta}\) until \(E(\vec{\theta})\) converges for the current orbital basis.

\small{\subsection*{Measurement of reduced density matrices}}
At VQE convergence for a given set of orbitals, the active-space one- and two-particle reduced density matrices (1-RDM and 2-RDM) are extracted from the optimized state 
$|\Psi(\vec{\theta}^{*})\rangle$. In second quantization, the spin-orbital RDMs are defined as
\begin{align}
\gamma_{pq}
&= \langle \Psi(\vec{\theta}^\star)|
\hat{a}_p^\dagger \hat{a}_q
|\Psi(\vec{\theta}^\star)\rangle,
\label{eq:rdm1-def} \\
\Gamma_{pqrs}
&= \langle\Psi(\vec{\theta}^\star)|
\hat{a}_p^\dagger \hat{a}_q^\dagger \hat{a}_s \hat{a}_r
|\Psi(\vec{\theta}^\star)\rangle.
\label{eq:rdm2-def}
\end{align}
Each operator $\hat{a}^{\dagger}_{p}\hat{a}_{q}$ or $\hat{a}^{\dagger}_{p}\hat{a}^{\dagger}_{q}\hat{a}_{s}\hat{a}_{r}$ is mapped to a short linear combination of Pauli strings under the chosen fermion–to–qubit transformation, so that Eq.~\eqref{eq:rdm1-def}–\eqref{eq:rdm2-def} reduce to expectation values of Pauli operators that are accessible on the quantum device. The RDM elements are assembled classically from these measurement results.

\small{\subsection*{Classical MC-PDFT energy from RDMs}}

Given the active-space RDMs, the MC-PDFT energy is evaluated on the classical processor. The total electronic energy in the active space is written as
\begin{equation}
\begin{split}
E_{\mathrm{MC\text{-}PDFT}}[\gamma,\Gamma]
&= \sum_{pq} h_{pq}\gamma_{pq} +
\frac{1}{4}\sum_{pqrs} h_{pqrs}\Gamma_{pqrs} \\
& + E_{\mathrm{ot}}\big[\rho(\gamma),\Pi(\gamma,\Gamma)\big],
\end{split}
\label{eq:mc-pdft-energy}
\end{equation}
where the first line collects the conventional multiconfigurational self-consistent field (MCSCF) energy contribution in the chosen orbital basis, and \(E_{\mathrm{ot}}\) denotes the on-top pair-density functional. The real-space electron density \(\rho(\mathbf{r})\) and on-top pair density \(\Pi(\mathbf{r})\) entering \(E_{\mathrm{ot}}\) are obtained by transforming the 1-RDM and 2-RDM to the atomic-orbital (AO) representation and contracting with atom-centered AO basis functions \(\{\chi_\mu(\mathbf{r})\}\), where \(\mu,\nu\) label AO basis functions and \(\mathbf{r}\) is the spatial coordinate. Using the spin-summed convention (i.e., \(\rho(\mathbf r)=\sum_\sigma \rho(\mathbf r,\sigma)\)), we write
\begin{align}
\rho(\mathbf{r})
&= \sum_{\mu\nu} D_{\mu\nu}\,\chi_\mu^{*}(\mathbf{r})\,\chi_\nu(\mathbf{r}), \\
\Pi(\mathbf{r})
&= \Pi^{(1)}(\mathbf{r}) + \Pi^{(2)}(\mathbf{r}),
\end{align}
where \(D_{\mu\nu}\) is the spin-summed AO 1-RDM obtained from \(\gamma_{pq}\) by a unitary transformation. The terms \(\Pi^{(1)}(\mathbf{r})\) and \(\Pi^{(2)}(\mathbf{r})\) denote the mean-field and cumulant contributions constructed from \(\gamma\) and \(\Gamma\) following standard MC-PDFT practice~\cite{li2014multiconfiguration}.

\small{\subsection*{Orbital-rotation parameterization and energy gradient}}
To incorporate the on-top functional variationally into the orbital optimization, we define its functional derivatives with respect to the RDMs:
\begin{equation}
v^{\mathrm{ot}}_{pq}
= \frac{\partial E_{\mathrm{ot}}}{\partial \gamma_{pq}},
\qquad
w^{\mathrm{ot}}_{pqrs}
= \frac{\partial E_{\mathrm{ot}}}{\partial \Gamma_{pqrs}}.
\label{eq:ot-derivs}
\end{equation}
The derivatives are evaluated via the chain rule from the explicit dependence of $E_{\text{ot}}[\rho(\gamma), \Pi(\gamma, \Gamma)]$ on $\rho$ and $\Pi$, together with the linear relationships $\rho(\gamma)$ and $\Pi(\gamma, \Gamma)$. Numerically, this step is performed on a real-space integration grid, in close analogy to the evaluation of exchange–correlation potentials in Kohn–Sham DFT, but using the on-top pair density instead of the spin densities.

To enable the variational optimization of molecular orbitals within the MC-PDFT framework, we employ a unitary transformation of the molecular orbital (MO) coefficient matrix:
\begin{equation}
    C' = Ce^{\boldsymbol{\kappa}}
\end{equation}
where $\boldsymbol{\kappa}$ is a real antisymmetric matrix, with elements $\kappa_{pq} = -\kappa_{qp},\: \kappa_{pq}\in\mathbb R$, parameterizing infinitesimal rotations between orbitals $p$ and $q$. This transformation allows us to adjust the shape of the molecular orbitals to minimize the total energy. The corresponding unitary operator in second quantization is:
\begin{equation}
    \hat{U}(\boldsymbol{\kappa}) = e^{-\hat{\kappa}}, \quad \text{with} \quad \hat{\kappa} = \sum_{p>q} \kappa_{pq}(\hat{E}_{pq} - \hat{E}_{qp})
\end{equation}
where $\hat{E}_{pq} = \sum_{\sigma} \hat{a}^{\dagger}_{p\sigma}\hat{a}_{q\sigma}$ are the electronic excitation operators. This sign convention reflects the standard passive versus active transformation: \(C' = C e^{\boldsymbol{\kappa}}\) updates the orbital basis, while the corresponding transformation acting on the many-electron wavefunction is generated by \(\hat{U}(\boldsymbol{\kappa}) = e^{-\hat{\kappa}}\). We denote this orbital-rotation unitary as \(\hat{U}_{\mathrm{orb}}\equiv \hat{U}(\boldsymbol{\kappa})\) in the following.

The key to optimizing the orbitals is to compute the energy gradient with respect to the orbital rotation parameters, $\kappa_{pq}$. Using the chain rule, the derivative of the MC-PDFT energy is:
\begin{equation}
\begin{split}
    \frac{\partial E_{\text{MC-PDFT}}}{\partial \kappa_{pq}} & = \sum_{rs} \frac{\partial E_{\text{MC-PDFT}}}{\partial \gamma_{rs}} \frac{\partial \gamma_{rs}}{\partial \kappa_{pq}} \\ & + \sum_{tuvw} \frac{\partial E_{\text{MC-PDFT}}}{\partial \Gamma_{tuvw}} \frac{\partial \Gamma_{tuvw}}{\partial \kappa_{pq}}
\end{split}
\end{equation}
where $\gamma_{rs}$ and $\Gamma_{tuvw}$ are the elements of the 1- and 2-RDMs, respectively. The orbital rotation gradient is defined as the derivative, 
\begin{equation}
    g_{pq} \equiv \frac{\partial E_{\text{MC-PDFT}}}{\partial \kappa_{pq}}
\end{equation}
Only the independent parameters with \(p>q\) are optimized; the full matrix form is obtained by enforcing antisymmetry.

In standard MCSCF theory, the orbital gradient can be expressed in terms of a generalized Fock matrix. In the context of MC-PDFT, we define an analogous MC-PDFT Fock matrix
\begin{equation}
    F_{pq}^{\text{MC-PDFT}} = F_{pq}^{\text{MCSCF}} + F_{pq}^{\text{ot}}
\end{equation}
where the latter term arises from the on-top functional. The orbital gradient then takes the form:
\begin{equation}
\begin{split}
    g_{pq} & = 2(F_{pq}^{\text{MC-PDFT}} - F_{qp}^{\text{MC-PDFT}}) \\ & = 2(F_{pq}^{\text{MCSCF}} + F_{pq}^{\text{ot}} - F_{qp}^{\text{MCSCF}} - F_{qp}^{\text{ot}})
\end{split}
\end{equation}
In practice, we use a quasi-Newton or steepest-descent algorithm to construct an updated orbital-rotation generator, $\boldsymbol{\kappa}^{(k)}$, from the gradient:
\begin{equation}
    \kappa^{(k)}_{pq} = -\alpha g^{(k)}_{pq}
\end{equation}
where $\alpha > 0$ is the step size, and the superscript $(k)$ denotes the iteration number. The many-electron orbital-rotation unitary at iteration $k$ is then obtained by exponentiation:
\begin{equation}
    \hat{U}_{\text{orb}}^{(k)} = e^{-\hat{\kappa}^{(k)}}
    \label{eq:orb-unitary}
\end{equation}

\small{\subsection*{Encoding the Orbital Rotation as a Quantum Circuit}}

The orbital-rotation generator \(\boldsymbol{\kappa}\) is computed classically from the orbital gradient and must be compiled into an executable quantum circuit that implements the corresponding many-electron unitary \(\hat{U}_{\mathrm{orb}}\).
This is achieved by viewing $\hat{U}_{\mathrm{orb}}$ as the time-evolution operator generated by an effective one-body Hamiltonian, $\hat{H}_{\mathrm{orb}}$. This Hamiltonian is constructed from the anti-Hermitian generator $\boldsymbol{\kappa}$ such that its evolution for a time $\Delta t$ reproduces the desired unitary:
\begin{equation}
\hat{U}_{\mathrm{orb}} = e^{-\hat{\kappa}} \equiv e^{-i\hat{H}_{\mathrm{orb}}\Delta t},
\quad
\hat{H}_{\mathrm{orb}} = \sum_{pq} K_{pq}\,\hat{E}_{pq},
\end{equation}
where the one-particle Hermitian matrix \(K\) is chosen such that \(K = -\frac{i}{\Delta t}\boldsymbol{\kappa}\) (with \(\boldsymbol{\kappa}\) real antisymmetric), ensuring that \(\hat{H}_{\mathrm{orb}}\) is Hermitian and \(\hat{U}_{\mathrm{orb}}\) is unitary.

To implement this on a quantum computer, the fermionic Hamiltonian $\hat{H}_{\mathrm{orb}}$ is first mapped to a qubit representation using the same fermion-to-qubit transformation chosen for the main electronic Hamiltonian. This results in a linear combination of Pauli strings:
\begin{equation}
    \hat{H}_{\mathrm{orb}} = \sum_{j} k_{j} \hat{P}_{j}
\end{equation}
Since the Pauli strings $\hat{P}_{j}$ do not mutually commute in general, a direct implementation of the corresponding unitary operator is intractable. Instead, we approximate the evolution using a Suzuki-Trotter~\cite{suzuki1976generalized} decomposition, which factorizes the operator into a product of simpler exponentials:
\begin{equation}
    \hat{U}_{\mathrm{orb}} = e^{-i\hat{H}_{\mathrm{orb}}\Delta t} \approx \prod_{j} e^{-i k_{j} \hat{P}_{j} \Delta t}.
\end{equation}
Each term $e^{-ik_{j} \hat{P}_{j} \Delta t}$ in this product can be efficiently compiled into a sequence of elementary one- and two-qubit gates. The complete sequence of gates forms a compiled quantum circuit $\hat{U}_{\mathrm{orb}}(\vec{\lambda})$, where the angles \(\vec{\lambda}\) are determined by the current one-particle generator \(\boldsymbol{\kappa}\) and are held fixed within each orbital-update iteration.

Finally, this orbital rotation circuit is appended to the original CASCI ansatz. The full quantum circuit for preparing the variationally improved trial state is thus given by:
\begin{equation}
    |\Psi_{\mathrm{final}}\rangle (\vec{\theta},\vec{\lambda}) = \hat{U}_{\mathrm{orb}}(\vec{\lambda}) U_{\mathrm{ansatz}}(\vec{\theta}) |\Psi_{\mathrm{HF}}\rangle.
\end{equation}
Operators act from right to left in the above expression. In this manner, the orbital-rotation unitary determined by the current generator \(\boldsymbol{\kappa}\) is systematically compiled into a quantum circuit and integrated into the variational state-preparation procedure.

\small{\subsection*{Self-consistent variational VQE-MC-PDFT procedure}}

The complete VQE-MC-PDFT algorithm proceeds as a self-consistent hybrid quantum-classical loop, designed to variationally optimize both the configuration-interaction coefficients and the molecular orbitals. The iterative procedure is as follows:

\begin{enumerate}
    \item \textbf{Initialization:} A classical Hartree--Fock (HF) calculation is performed to obtain an initial molecular orbital basis. The corresponding HF determinant is prepared as the reference state $|\Psi_{\mathrm{HF}}\rangle$ on the quantum device.

    \item \textbf{VQE for CI Coefficients:} For the current fixed orbital basis, a VQE optimization of the ansatz parameters $\vec{\theta}$ in $U_{\mathrm{ansatz}}(\vec{\theta})$ is performed to find the optimal multi-configurational wavefunction. Upon convergence, the active-space 1-RDM and 2-RDM are computed by measuring the quantum state.

    \item \textbf{Classical MC-PDFT Analysis:} The measured RDMs are passed to a classical computer. This classical module evaluates the on-top functional $E_{\mathrm{ot}}$, computes its derivatives $v^{\mathrm{ot}}$ and $w^{\mathrm{ot}}$, constructs the generalized Fock matrix $F_{pq}^{\mathrm{MC-PDFT}}$, and determines the orbital gradient $g_{pq}$.

    \item \textbf{Orbital Rotation Update:} The orbital gradient is checked for convergence. If it is above a defined threshold, a new orbital rotation generator $\boldsymbol{\kappa}$ is constructed from $g_{pq}$ and mapped to its second-quantized form \(\hat{\kappa}\). 
    This generator is exponentiated to define the many-electron unitary \(\hat{U}_{\mathrm{orb}} = e^{-\hat{\kappa}}\).

    \item \textbf{Circuit Encoding and Iteration:} The many-electron unitary $\hat{U}_{\mathrm{orb}}$ is compiled into a quantum circuit \(\hat{U}_{\mathrm{orb}}(\vec{\lambda})\) via Trotterization. The state-preparation unitary for the next cycle is updated to the combined circuit $\hat{U}_{\mathrm{orb}}(\vec{\lambda})U_{\mathrm{ansatz}}(\vec{\theta})$. 
    The algorithm then returns to step 2 for the next self-consistent iteration, where \(\vec{\theta}\) is re-optimized for the updated circuit while \(\vec{\lambda}\) is fixed by the current \(\boldsymbol{\kappa}\).

\end{enumerate}

This procedure is repeated until both the total energy and the norm of the orbital gradient converge. Through this seamless integration of quantum state preparation, classical functional evaluation, and quantum-implemented orbital rotations, the VQE--MC-PDFT algorithm achieves a fully variational optimization of the electronic wavefunction on near-term quantum hardware.

\small{\subsection*{Error mitigation}}
The fidelity of the variational quantum eigensolver is critically dependent on the precise estimation of the energy expectation value, a quantity derived from the probability distribution of measurement outcomes. In any physical implementation, this distribution is inevitably corrupted by noise, with measurement readout error being a particularly dominant source. This error manifests as a classical stochastic process that distorts the ideal probability distribution, potentially leading the VQE optimization to converge on non-physical minima. Therefore, the application of a robust error mitigation protocol is an essential step for achieving chemically meaningful results.

The standard approach~\cite{maciejewski2020mitigation} models the measurement error channel with a linear calibration matrix, $\Lambda$, which maps the ideal probability vector of measurement outcomes, $\mathbf{b}_{\text{ideal}}$, to the experimentally observed noisy vector, $\mathbf{b}_{\text{noisy}}$:
\begin{equation}
    \mathbf{b}_{\text{noisy}} = \Lambda \mathbf{b}_{\text{ideal}}
\end{equation}
For an $N$-qubit system, the characterization of the full $2^N \times 2^N$ matrix $\Lambda$ is exponentially costly. A common and practical simplification assumes that measurement errors are local and uncorrelated. This allows the total calibration matrix to be approximated as a tensor product of single-qubit calibration matrices:
\begin{equation}
    \Lambda \approx \bigotimes_{j=1}^{N} \Lambda_{j}
\end{equation}
Each single-qubit matrix $\Lambda_j$ is a $2 \times 2$ stochastic matrix that characterizes the readout fidelity on the $j$-th qubit:
\begin{equation}
    \Lambda_j = \begin{pmatrix}
 p(0|0) & p(0|1) \\
 p(1|0) & p(1|1)
\end{pmatrix} =: \begin{pmatrix}
 1-\epsilon_j & \gamma_j \\
 \epsilon_j & 1 - \gamma_j
\end{pmatrix}
\end{equation}
where $p(x|y)$ is the probability of measuring state $|x\rangle$ given preparation in state $|y\rangle$. The parameters $\epsilon_j$ and $\gamma_j$ are calibrated empirically by preparing a set of known product states $\mathcal{T}$ and measuring the resulting outcome frequencies $m(y, x)$:
\begin{align}
    \epsilon_j & = \frac{\sum_{x, y}{m(y, x) \langle 1 | y_j \rangle \langle x_j | 0 \rangle}}{\sum_{x, y} m(y, x) \langle x_j | 0 \rangle} \\
    \gamma_j & = \frac{\sum_{x, y}{m(y, x) \langle 0 | y_j \rangle \langle x_j | 1 \rangle}}{\sum_{x, y} m(y, x) \langle x_j | 1 \rangle}
\end{align}
While this tensor-product model offers a tractable first-order correction, its core assumption of qubit independence is a significant oversimplification. It fundamentally neglects the effects of crosstalk and state-dependent noise, where the measurement of one qubit is correlated with the state or measurement of its neighbors.

To address these deficiencies and achieve a more faithful error characterization, our methodology adopts a sophisticated framework inspired by the Finite Element Method (FEM)~\cite{tan2024qufem}. This approach replaces the single, monolithic correction with an iterative, divide-and-conquer strategy that can model local correlations. The central idea is to partition the system of $N$ qubits into smaller, computationally manageable clusters and to model the noise locally within these groups using sub-noise matrices. The mitigation is then applied as a sequence of refinements, where each step corrects the probability distribution from the previous one. Let $\mathbf{b}^{(0)}$ be the initial raw, noisy probability vector. The iterative correction is formally described as:
\begin{equation}
    \mathbf{b}^{(i+1)} = \left( M_{i,1} \otimes M_{i,2} \otimes \dots \otimes M_{i,K_i} \right)^{-1} \mathbf{b}^{(i)}
\end{equation}
In each iteration $i$, the qubits are partitioned into a set of clusters $G_i = \{g_{i,1}, \dots, g_{i,K_i}\}$, and a corresponding set of sub-noise matrices $\{M_{i,j}\}$ is applied. By systematically varying the qubit partitioning scheme across iterations, this method can probe different local correlation structures, progressively purifying the probability distribution. A key advantage of this framework is its dynamic and input-specific nature; the sub-noise matrices are generated conditionally, taking into account the specific context of the circuit being measured, which allows the model to capture complex error dependencies that a global calibration matrix would miss. This input-specific character is the mechanism by which we leverage electronic density information for correction; the sub-noise matrices are conditioned on the measured probability distribution of the quantum state, implicitly using information about the system's electronic density to build a more physically realistic and accurate noise model than a generic, state-independent calibration could provide.

While this sophisticated framework corrects for errors at the measurement stage, a comprehensive strategy must also contend with the accumulation of noise during the circuit execution. We employ two prominent paradigms for mitigating these gate errors: Zero-Noise Extrapolation (ZNE)~\cite{giurgica2020digital} and Clifford Fitting (CF)~\cite{czarnik2021error}. ZNE is a prophylactic approach that infers the ideal, zero-noise expectation value by systematically amplifying the noise in a controlled manner and extrapolating the results back to the zero-noise limit. The canonical implementation for this noise amplification is *local unitary folding*, where each gate $G$ is replaced by a sequence $G(G^\dagger G)^k$. This amplifies the associated gate noise by a factor of approximately $\lambda = 2k+1$. By measuring the expectation value $E(\lambda_i)$ for a set of noise levels $\{\lambda_i\}$, we fit a noise model, such as a polynomial $E(\lambda) = E_0 + a_1\lambda + \dots$, to the data. The mitigated expectation value $E_0$ is then obtained by evaluating the fitted function at the zero-noise limit, $\lambda=0$. As an alternative, data-driven paradigm, Clifford Fitting leverages the unique properties of the Clifford group—a set of quantum operations that can be efficiently simulated classically. The core principle of CF is to learn a function that maps a noisy expectation value to its ideal, error-free counterpart, typically using a linear model: $E_{\text{ideal}} = f_{\text{CF}}(E_{\text{noisy}}) = a E_{\text{noisy}} + b$. To learn the parameters $a$ and $b$, a training set of quantum circuits is constructed by replacing the non-Clifford gates in the original variational ansatz with randomly chosen Clifford gates. For each of these nearly-Clifford circuits, the ideal expectation value can be computed classically, while the noisy value is obtained from execution on the quantum hardware. A regression performed on these classical-quantum data pairs yields the error mitigation function $f_{\text{CF}}$, which can then be applied to the noisy expectation value obtained from the actual VQE experiment.

By combining FEM‑inspired readout correction with ZNE or CF, we obtain a calibration tailored to the measured circuit distribution.

This structured methodology provides a scalable path to mitigating complex, correlated readout errors, producing a corrected probability distribution that more accurately represents the ideal quantum state. This high-fidelity distribution is then used to compute the final expectation values required for the VQE-MC-PDFT algorithm, forming a critical step in achieving chemically accurate results on noisy quantum hardware.

\section*{Data availability}\label{da}
The data that support the findings of this study are available within the Article and its Supplementary Information. Specifically, the source data underlying the potential energy curves and excitation energies presented in Figs.~\ref{ground-c2}--\ref{excited-benzene} and Table~\ref{tab:c2_spectroscopy} are provided in the Supplementary Information. Any additional data are available from the corresponding authors upon reasonable request.

\section*{Code availability}\label{ca}
The computer codes used to generate the numerical results and execute the VQE-MC-PDFT workflows in this study are available on GitHub at \url{https://github.com/clzoc/VQE-MC-PDFT}.

\section*{Acknowledgments}
Y.D. was supported by the National Key R\&D Program of China (Grant No. 2023YFA1009403), the National Natural Science Foundation of China (Grant No. 62472175), the “Digital Silk Road” Shanghai International Joint Lab of Trustworthy Intelligent Software (Grant No. 22510750100), and the Fundamental Research Funds for the Central Universities.
X.H. was supported by the Shanghai Municipal Science and Technology Commission (Grant No. 25511102400), the National Natural Science Foundation of China (Grant Nos. 92477103 and 22273023), the Shanghai Municipal Natural Science Foundation (Grant No. 23ZR1418200), the Shanghai Frontiers Science Center of Molecule Intelligent Syntheses, and the Fundamental Research Funds for the Central Universities. We also acknowledge the Supercomputer Center of East China Normal University (ECNU Multifunctional Platform for Innovation 001) for providing computing resources.

\section*{Author Contributions}
Z.L. and Y.D. conceived the research. Z.L. implemented the VQE-MCPDFT algorithm and conducted the numerical simulations. Y.C. and Y.M. assisted in the computational benchmarks and data analysis. X.H. and Y.D. supervised the project and co-wrote the manuscript. All authors discussed the results and commented on the final version of the manuscript.

\section*{Competing Interests}
The authors declare no competing interests.


\end{document}